\documentclass[reprint,aps,amsmath,amssymb]{revtex4-2}
\usepackage{color}
\usepackage{graphicx}
\usepackage{dcolumn}
\usepackage{bm}
\usepackage[colorlinks=true,citecolor=blue,urlcolor=blue,linkcolor=blue,hyperindex]{hyperref}
\usepackage{mathtools}
\usepackage{float}
\usepackage{booktabs}
\usepackage[caption=false,justification=justified]{subfig}
\makeatother
\begin{document}

\preprint{APS/123-QED}

\title{Superband: an Electronic-band and Fermi surface structure database of superconductors}


\author{Tengdong Zhang$^{1}$}
\author{Chenyu Suo$^{1}$}
\author{Yanling Wu$^{1}$}
\author{Xiaodan Xu$^{1}$}
\author{Yong Liu$^{1}$}
\email{yongliu@ysu.edu.cn}
\author{Dao-Xin Yao$^{2}$}
\email{yaodaox@mail.sysu.edu.cn}
\author{Jun Li$^{1}$}
\email{ljcj007@ysu.edu.cn}
\affiliation{
$^1$Key Laboratory for Microstructural Material Physics of Hebei Province, School of Science, Yanshan University, Qinhuangdao 066004, China.\\
$^2$State Key Laboratory of Optoelectronic Materials and Technologies, Guangdong Provincial Key Laboratory of Magnetoelectric Physics and Devices, School of Physics, Sun Yat-Sen University, Guangzhou 510275, Peoples Republic of China.}%

\begin{abstract}
In comparison to simpler data such as chemical formulas and lattice structures, electronic band structure data provide a more fundamental and intuitive insight into superconducting phenomena. In this work, we generate superconductor's lattice structure files optimized for density functional theory (DFT) calculations. Through DFT, we obtain electronic band superconductors, including band structures, density of states (DOS), and Fermi surface data. Additionally, we outline efficient methodologies for acquiring structure data, establish high-throughput DFT computational protocols, and introduce tools for extracting this data from large-scale DFT calculations. As an example, we have curated a dataset containing information on 2474 superconductors along with their experimentally determined superconducting transition temperatures ($T_c$), which is well-suited for machine learning applications. This work also provides guidelines for accessing and utilizing this dataset. Furthermore, we present a neural network model designed for training with this data. All the aforementioned data and code are publicly available at \href{http://www.superband.work/}{http://www.superband.work/}.

\end{abstract}
\maketitle

\section{\label{sec:intro}Introduction}

\begin{figure}[b]
	\centering
	\includegraphics[width=0.35\textwidth]{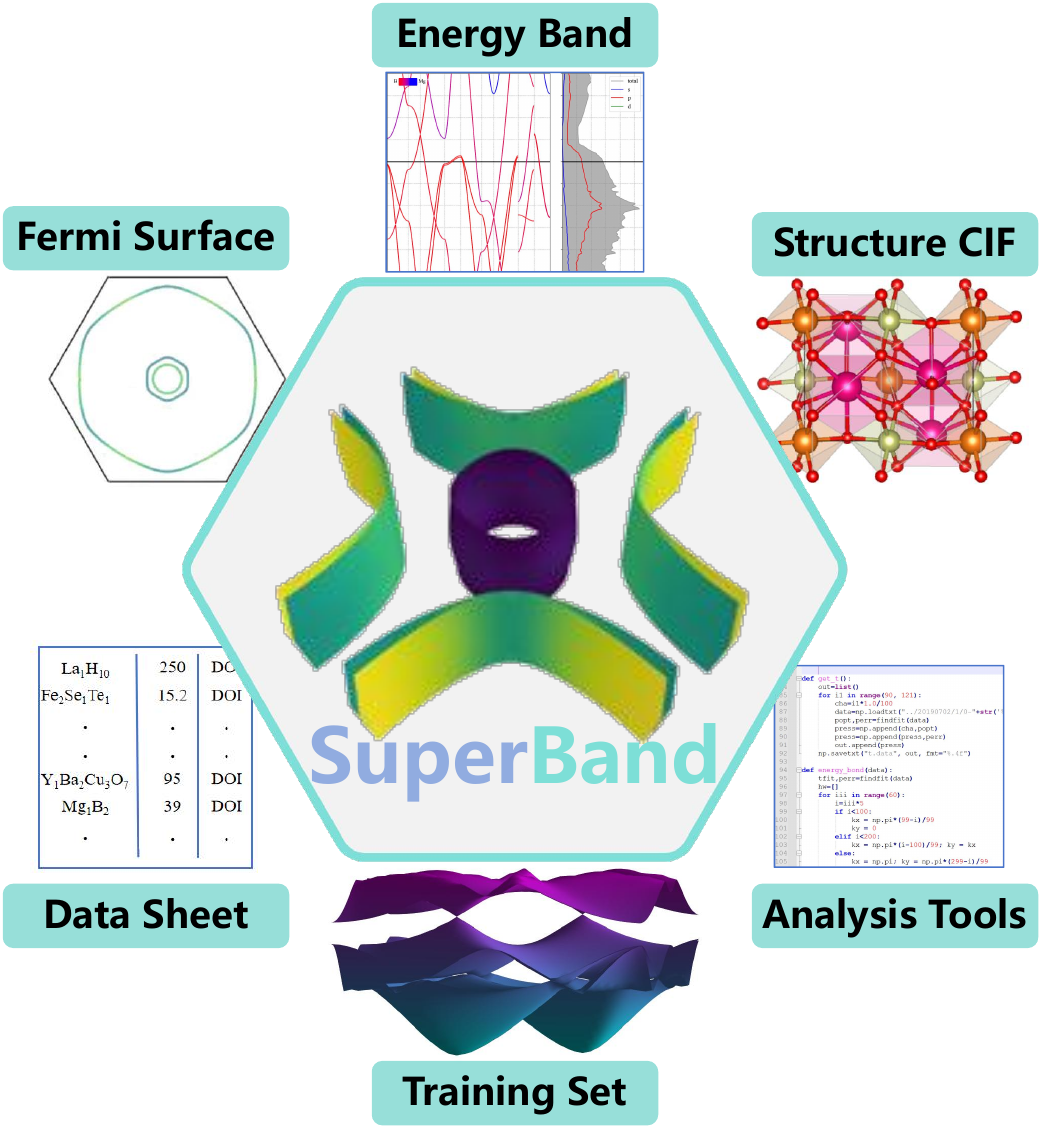}
	\caption{\label{fig1}Superband: an electronic-band and Fermi surface structure database of superconductors}
\end{figure}

\begin{figure*}[t]
	\centering
	\includegraphics[width=0.65\textwidth]{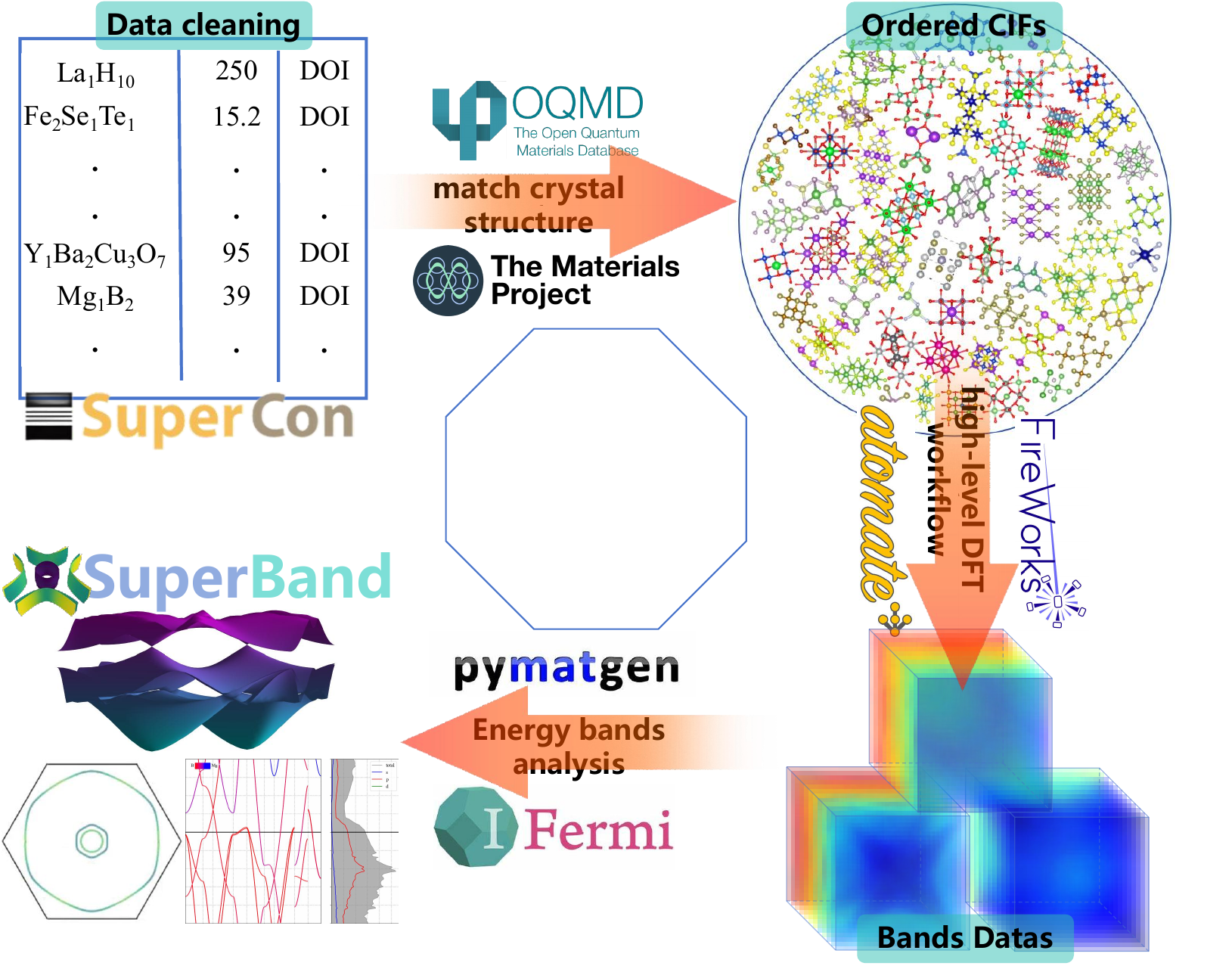}
	\caption{\label{fig2}Computational and data cleaning details of Superband. Chemical formulas and $T_c$ data are sourced from the SuperCon database \cite{m900p111d}, while structural CIF data are obtained from the Materials Project (MP) \cite{MPS} and the Open Quantum Materials Database (OQMD) \cite{Kirklin2015}. High-throughput calculations for these materials are conducted using the Atomate open-source package \cite{Mathew2017,Jain2011}. The FireWorks package \cite{CPE:CPE3505} is utilized for managing the DFT workflow, with Pymatgen \cite{Jain2013,Ong2015} and Ifermi \cite{Ganose2021} employed to extract energy band data.
 }
\end{figure*}
The phenomenon of zero electrical resistance in a material is of profound scientific and practical significance, referred to as superconductivity. This unique state allows electric current to flow through a material without any energy dissipation, making it an essential field of study with numerous potential applications. However, the practical use of superconductors is often constrained by the requirement for extremely low temperatures or high pressures. Since its initial discovery in 1911, the quest for superconductors that function at higher temperatures has been a major focus, as such advancements would enable a wider range of technological applications.

Superconductivity is a well-documented phenomenon, with over 10,000 superconductors identified to date \cite{m900p111d,Sommer2023}. Prominent examples include cuprate \cite{Bednorz1986}, iron-based  \cite{Stewart2011} and nickel-based superconductors\cite{Sun2023}, which highlight the typical progression in the field: experimental physicists first synthesize new superconductors, followed by theoretical physicists who seek to unravel the fundamental mechanisms of superconductivity through a variety of models and theoretical frameworks. Despite the existence of numerous theories in condensed matter physics that attempt to explain superconductivity, predicting new high-temperature superconductors remains one of the greatest challenges in the field.

In condensed matter theory, energy-band theory serves as a cornerstone for understanding the electronic properties of materials. First-principles calculations based on density functional theory (DFT) play a crucial role in this regard, offering detailed insights into a material’s electronic band structure and density of states (DOS). These elements are instrumental in determining the electrical properties of a material \cite{Martin2004}. Since superconductivity is inherently an electrical property, it follows that the energy-band theory derived from DFT should be applicable for explaining and predicting superconducting behavior \cite{Lueders2005}. 

Theoretically, the electronic band structure obtained from DFT calculations provides essential parameters for understanding superconducting behavior. These parameters are critical for elucidating both conventional superconductors, such as those explained by BCS theory (e.g., the superconducting gap and electron-phonon coupling constants \cite{Giustino2017}), and unconventional superconductors, where strong correlations \cite{Aichhorn2010} and spin fluctuations \cite{Graser2010} play a pivotal role.

For instance, the electron-phonon coupling constant in ambient-pressure BCS superconductors like MgB$_2$, which boasts a relatively high transition temperature \cite{Bohnen2001}, can be extracted through DFT calculations. Likewise, DFT has been instrumental in identifying the key parameters in high-pressure hydrogen-rich superconductors LaH$_{10}$ \cite{Drozdov2019}. Furthermore, two-dimensional carbon-based materials \cite{Li2022,Si2013}, nonlinear phonon properties in YBa$_2$Cu$3$O${6.5}$ \cite{Mankowsky2014}, and magnetic interactions in iron-based superconductors \cite{Graser2010} are examples where DFT has significantly contributed to understanding unconventional superconductivity.

Moreover, DFT has provided insights into the tight-binding model parameters \cite{Cao2008}, electronic Coulomb correlation terms in iron-based superconductors \cite{Aichhorn2010}, spin-orbit coupling in heavy fermion systems \cite{Samokhin2004}, and interlayer interactions in bilayer twisted graphene \cite{Carr2018}. It also helps illuminate the  $\sigma$-bonds in high-pressure nickel-based superconductors \cite{Luo2023}, the superconducting pairing symmetry in bilayer silicene \cite{Liu_2013}, and the unconventional pairing mechanisms in two-dimensional carbon materials \cite{Li2020,Ye2023}. These examples showcase the breadth of factors that DFT can address in advancing our understanding of both conventional and unconventional superconductivity.

In contrast to simpler data, such as chemical formulas and lattice structures, electronic band structure data provides a more fundamental and intuitive perspective on superconducting phenomena. This deeper insight is particularly relevant in the context of recent advancements in big data processing techniques, including machine learning (ML) approaches \cite{Carleo2019}. The potential of ML to analyze complex electronic properties highlights the need for a comprehensive database of electronic band structures. Such a database would enable large-scale analyses, fostering the discovery of new superconductors and enhancing the understanding of their underlying mechanisms. The development of this resource is essential for advancing both theoretical and experimental research in the field of superconductivity.

In this paper, we introduce Superband, a comprehensive electronic band and Fermi surface structure database for superconductors, as depicted in Fig. \ref{fig1}. We generate lattice structure files optimized for DFT calculations and, through these calculations, obtain crucial electronic band data for experimentally realized superconductors. This dataset includes the electronic band structure, DOS, and Fermi surface information. Additionally, we outline methods for the efficient acquisition of structural data, provide high-throughput DFT calculation protocols, and offer programs designed to extract the aforementioned data from large-scale DFT computations.

To demonstrate the utility of this database, we have compiled a dataset of 2,472 superconductors, including their experimentally determined superconducting transition temperatures ($T_c$), which is ideal for ML applications. Detailed instructions on accessing this dataset are provided in this work. Furthermore, we introduce a neural network model capable of training on this dataset, enabling predictions of superconducting properties based on electronic structure data. All data, protocols, and neural network models discussed in this work are publicly available at: \href{http://www.superband.work/}{http://www.superband.work/}.

\section{\label{sec:method}Computational Details}

The chemical formulas and $T_c$ data for superconductors presented in this paper are sourced from the 2023 edition of the SuperCon database \cite{m900p111d}. This extensive database contains information on 33,458 materials, including 7,190 non-superconducting compounds and 26,268 superconductors with experimentally measured $T_c$ values. To ensure the most up-to-date dataset, we supplemented these materials with superconductors newly identified after 2022 by reviewing publicly available literature.

As depicted in Fig. \ref{fig2}, the crystal structure data utilized in this study are primarily obtained from the Materials Project \href{https://next-gen.materialsproject.org/}{https://next-gen.materialsproject.org/} \cite{MPS}, with additional contributions from the Open Quantum Materials Database (OQMD) \href{https://www.oqmd.org/}{https://www.oqmd.org/} \cite{Kirklin2015}. Since a significant proportion of superconductors are derived through doping parent compounds with various elements, we adopt the 3DSC methodology \cite{Sommer2023} for lattice doping. To handle the complexities of doped structures, supercell processing is applied, replacing doped atoms and generating ordered crystal lattice data files compatible with density functional theory (DFT) calculations. 

The projector-augmented wave (PAW) method, implemented in the Vienna Ab initio Simulation Package (VASP), is employed to carry out our DFT calculations \cite{Kresse1993}. High-throughput DFT calculations are facilitated by the Atomate open-source package \cite{Mathew2017}, with parameter settings derived from the MIT High-Throughput Project \cite{Jain2011}. For workflow automation, we employ the FireWorks package \cite{CPE:CPE3505}, which efficiently manages the task flow for structure optimization, static calculations, self-consistent field (SCF) calculations, and band structure determinations at high-symmetry points in reciprocal space.

To extract relevant data for analysis, including band structure and DOS, we utilize the Pymatgen package \cite{Jain2013,Ong2015}. For Fermi surface generation, analysis, and visualization, the Ifermi package \cite{Ganose2021} is used, enabling detailed examination of electronic properties crucial for understanding superconducting mechanisms.

\section{\label{sec:sc}Data Cleaning}

\begin{figure}[]
	\centering
	\includegraphics[width=0.45\textwidth]{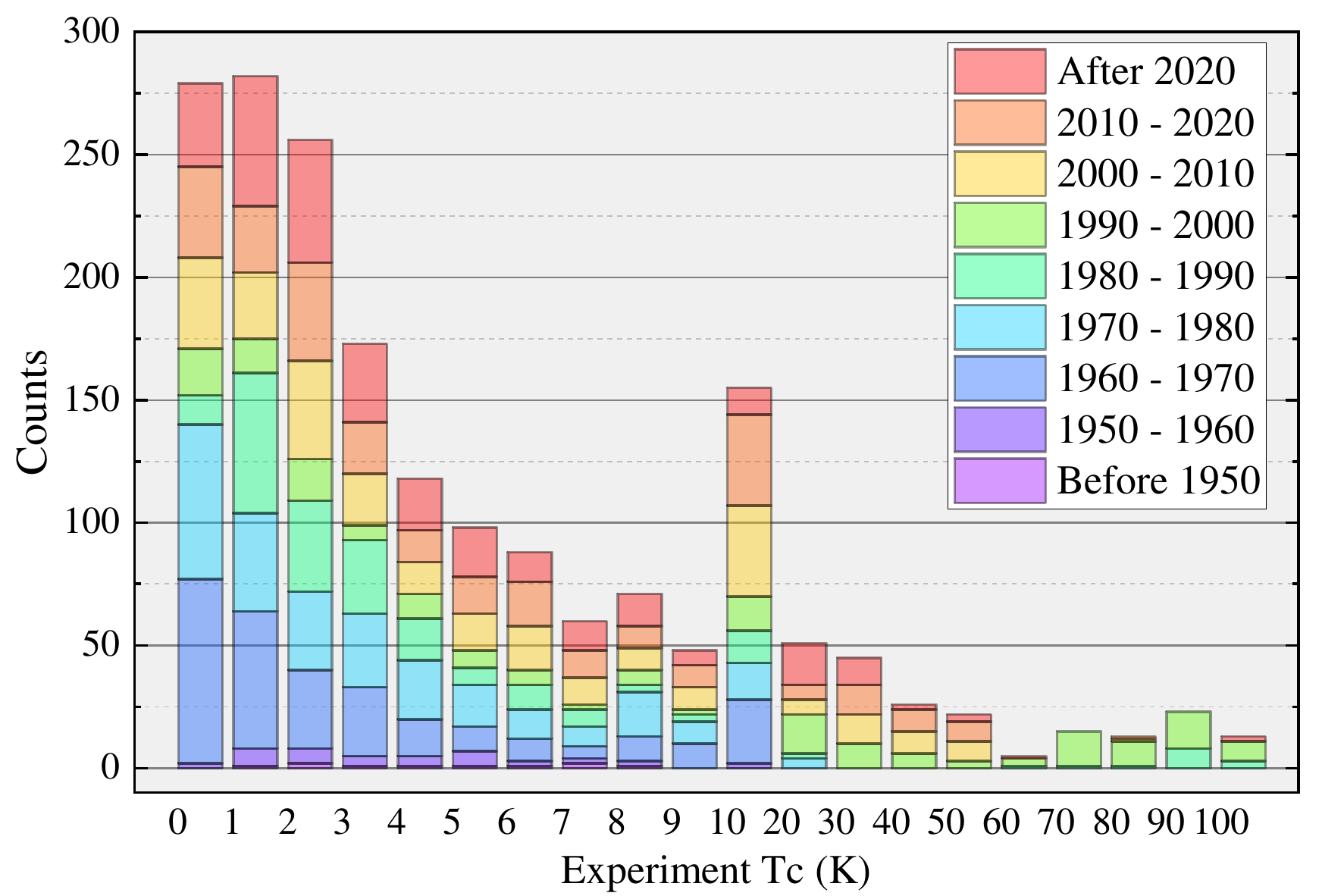}
	\includegraphics[width=0.45\textwidth]{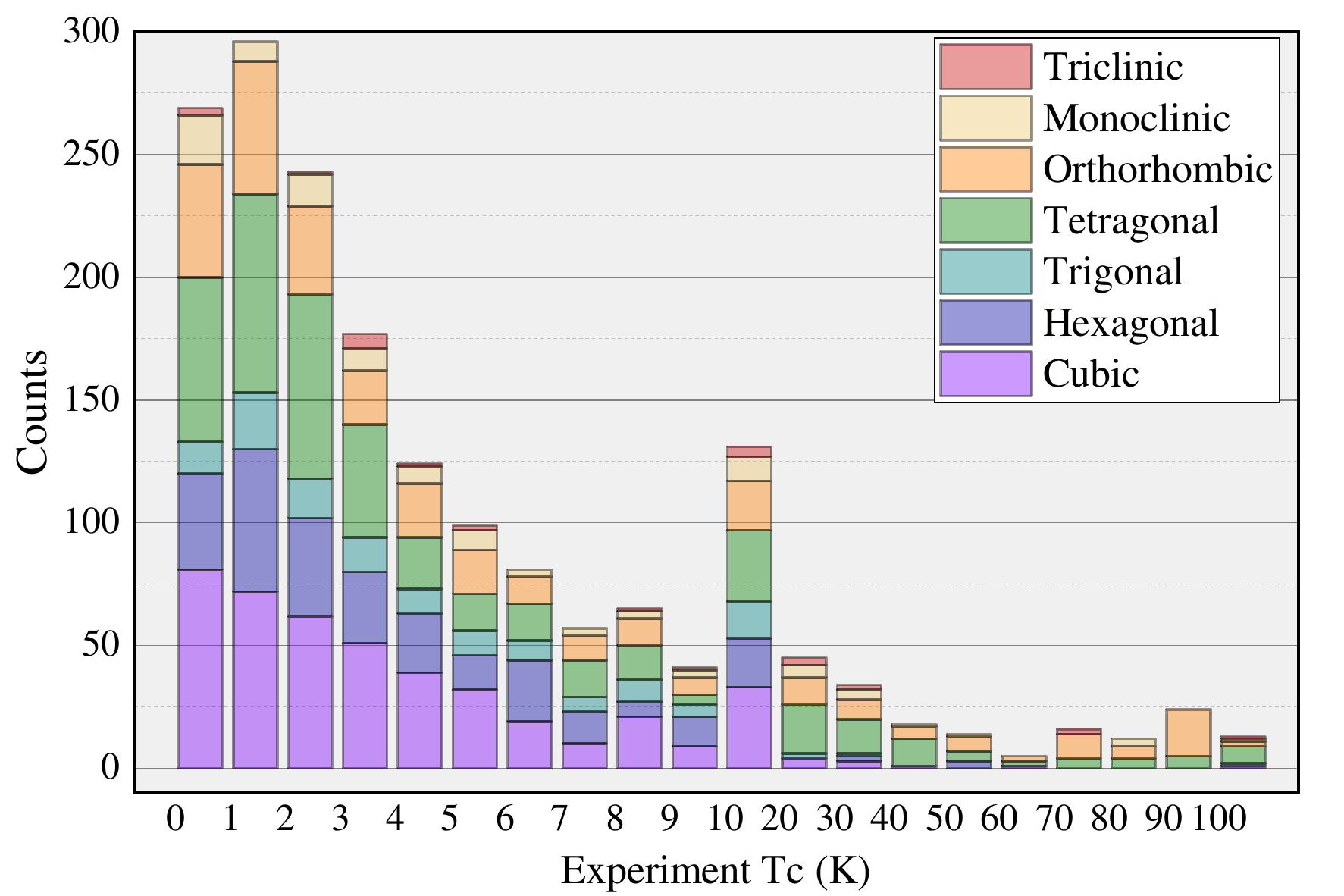}
	\caption{\label{cif4}Year distribution (Top) and crystal system distribution (bottom) of superconductors in Superband.}
\end{figure}
The SuperCon database \cite{m900p111d} contains numerous duplicate data entries, necessitating a rigorous data cleaning process. A key distinction of this work, compared to previous studies, lies in the determination of ordered crystal lattices for superconductors suitable for DFT calculations. The initial phase involved retrieving crystallographic information files (CIFs) for lattice structures from relevant databases, including the MP and the OQMD. It should be noted that CIF files obtained from these public sources often contain disordered structures. In cases where CIF files were unavailable, we construct some disordered lattice structure files manually.

To address this, we employ an order transformation method that retains only ordered structures with the lowest Ewald energy \cite{Ong2015}. This method efficiently standardizes lattice structures with co-occupying atoms to generate ordered configurations. However, the method encounters difficulties when applied to materials with multiple-element co-occupations or a large number of co-occupied atomic sites. Consequently, we retained 14 materials for which disorder could not be resolved, including K$_2$RbC$_{60}$ (ID 15960), TiVNbTa (ID 16063), and Cu$_{0.65}$La$_{1.83}$Ni$_{0.35}$Sr$_{0.17}$O$_4$ (ID 17788). These materials were excluded from further DFT calculations due to unresolved structural complexities.

Subsequently, we applied the 3DSC methodology \cite{Sommer2023} to handle chemical formulas, including the definitions for exact matching, similarities, doping, and unmatched cases. This methodology is applied to the SuperCon database \cite{m900p111d} to determine whether the chemical formulas could be matched with the ordered structures collected. For materials in the SuperCon database accompanied by space group information, a space group matching analysis is also performed on the relevant CIF files to identify the most closely corresponding material structure.

When a fully matching or similar CIF file could not be identified, we search for materials with chemically doped formulas. If the doping concentration exceeded 0.75, the doped atoms are replaced. For doping concentrations exceeding 0.45 (0.29, 0.19, 0.1), supercell expansions of $1\times1\times2$ ($1\times1\times3$, $2\times2\times1$, $2\times2\times2$) are performed to accommodate the doped atoms. The doped atoms are then replaced while preserving the lattice symmetry as much as possible. This process is repeated until the expanded and substituted supercell achieve chemical similarity with the given chemical formulas.

It is important to note that the introduction of doping does not necessarily alter the $T_c$ of a material. In some cases, the incorporation of dopants has little to no discernible effect on the superconducting properties. For such doped superconductors, it is sufficient to disregard minor dopants that do not significantly impact superconductivity, as seen in SiV$_3$-based superconductors. A threshold of 0.2 is thus established to differentiate between doping and similarity for these materials.

However, for certain other systems, such as iron-based superconductors, even a small amount of elemental doping can substantially shift the Fermi level or modify DOS near the Fermi surface. These changes can markedly enhance or suppress superconductivity, often accompanied by a significant shift in $T_c$. For such materials, a more stringent threshold of 0.1 is applied to distinguish between doping and similarity, given the pronounced sensitivity of their superconducting properties to minor doping modifications.
\begin{figure}
\centering
\includegraphics[width=0.45\textwidth]{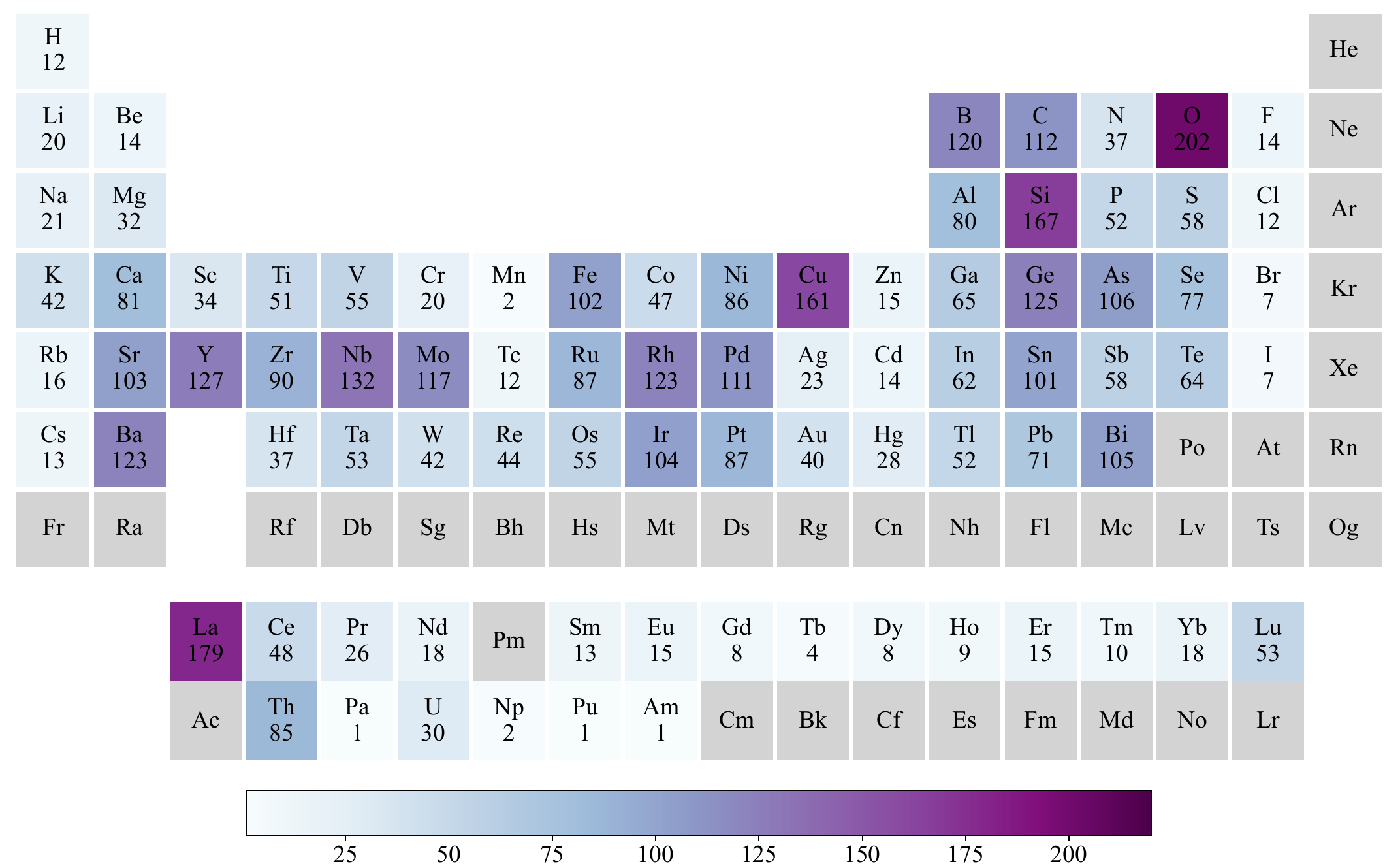}
\caption{\label{cif3}The elemental distribution of superconductors found experimentally. For any superconductor containing element A, the count for element A is incremented by 1. By following this process, we obtain the elemental distribution across all superconductors\cite{Li2024}. }
\end{figure}
\begin{figure*}[t]
	\centering
	\includegraphics[width=0.65\textwidth]{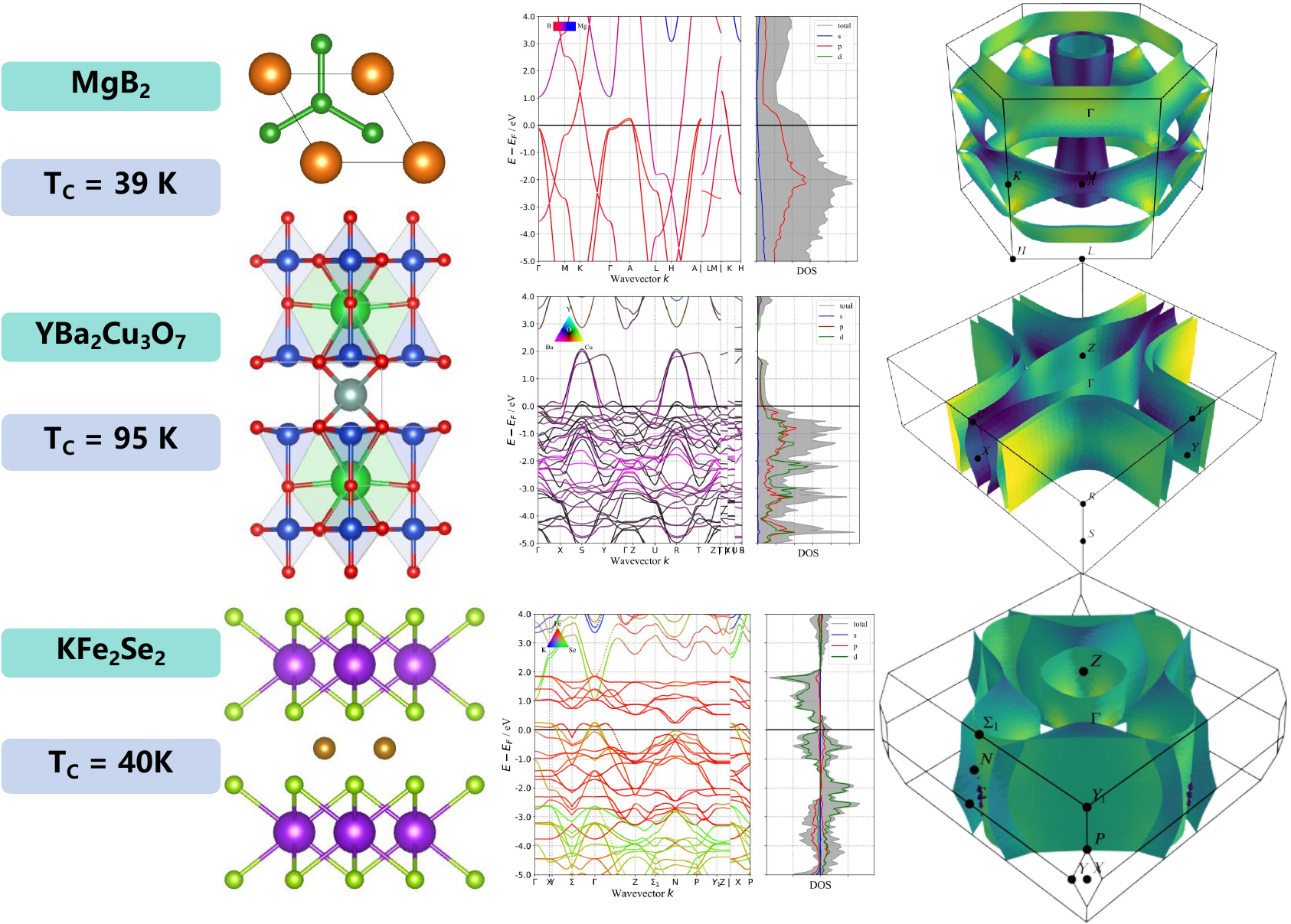}
	\caption{\label{cif5} Typical band data of three superconductors in Superband\cite{Bohnen2001,Collocott1988,Ying2012}. For each material, we provide space group, crystal structure, electronic band structure, and Fermi surface. }
\end{figure*}

Following the matching process with CIF files, we obtained data for 8,590 materials with non-duplicate chemical formulas, including 6,780 superconductors. Notably, compared to the reports on superconductors, there is a significant scarcity of reports on non-superconducting materials. Although the number of non-superconducting materials likely far exceeds that of superconductors, research on superconductivity often omits such materials from published studies.

In the realm of ML and big data research, this lack of data on non-superconducting materials hinders the reliability of predictions related to superconductivity. Non-superconducting materials are just as critical to the study of superconductivity, as they offer valuable insight into the boundaries of superconducting behavior. Therefore, we also provide data for 1,780 materials that have been experimentally verified to lack superconducting properties.

Notably, a significant portion of the 6,780 superconductors are represented by the same CIF file. We identified a total of 1,763 unique CIF files. It is inappropriate to classify a material as a distinct superconductor based on a minor doping of 0.01 of another element. As a result, the CIF file itself is used as the definitive criterion for identifying unique superconductors in this study. Therefore, the subsequent sections of this paper focus exclusively on the 1,763 superconductors corresponding to these unique CIF files, as depicted in Fig. \ref{cif3}.

A comprehensive summary of the literature documenting the initial experimental synthesis of these superconductors is provided. For 159 materials, no corresponding references were found. However, for the remaining 1,604 superconductors, relevant publications are identified. As illustrated in Fig. \ref{cif4}(a), the proportion of superconductors with $T_c$ below 30 K remains consistent across various periods, suggesting that the discovery of new superconductors is largely stochastic. Additionally, the distribution of superconductors relative to their $T_c$ follows an inverse relationship, except for those with $T_c$ $<$ 2 K. The 1970s saw the advent of superconductors with $T_c$ $>$ 30 K, most notably with the discovery of cuprate superconductors, which triggered a surge in high-temperature superconductor research during the 1980s.

The use of CIF files enables precise characterization of material properties via Pymatgen's structure tool, as shown in Fig. \ref{cif4}(b). Among superconductors, the most prevalent crystalline structure is tetragonal, which appears in 453 distinct cases. This is followed closely by cubic symmetry in 439 cases, with the fewest occurrences noted for monoclinic (112 instances) and triclinic (27 instances). The tendency of superconductors to favor high-symmetry structures aligns with Matthias' hypothesis regarding the correlation between symmetry and superconductivity. However, for materials with $T_c$ $>$ 10 K, a significant decline in the proportion of cubic superconductors is observed, coinciding with a marked increase in orthorhombic superconductors, which exhibit lower symmetry.

For superconductors with $T_c$ values greater than 40 K, the majority of unconventional superconductors that surpass the McMillan limit tend to have either tetragonal or orthorhombic symmetry. This shift suggests that structures with lower symmetry may play a key role in high-temperature superconductivity, especially in systems where conventional electron-phonon interactions are insufficient to explain the observed $T_c$.

\section{\label{sec:con}Band data}

The availability and standardization of data are critical prerequisites for the development of ML models aimed at predicting material properties. In our DFT calculations, the electronic bands of different materials show significant variations due to the MIT-initialized DFT parameter settings \cite{Jain2011}. Initially, lattice symmetry is considered to reduce computational costs, but the equivalent k-point values differ across space groups. Moreover, the k-space mesh density must be adjusted based on the number of atoms and lattice dimensions in each unit cell to enhance the accuracy of the calculations.

To address the normalization of k-space band data, we employ IFermi package \cite{Ganose2021} to standardize the k-space by considering only symmetry-equivalent k-points. Following this, interpolation techniques are applied to standardize the k-space mesh coordinates onto a uniform k grid of 32 $\times$ 32 $\times$ 32, ensuring consistency across various materials for ML applications.

After completing the standardization process, the number of electronic bands varies among different materials. In constructing a standardized dataset for ML, one could theoretically pad the training set tensors with zero tensors to maintain uniformity. However, this approach wastes computational resources and diminishes the efficiency of the calculations. Studies on both conventional and unconventional superconductors have demonstrated that the DOS near the Fermi surface has a substantial impact on superconductivity, while bands far from the Fermi surface contribute minimally \cite{Li2020,Ye2023}. Therefore, focusing on the electronic bands in close proximity to the Fermi surface is more computationally efficient and enhances the relevance of the dataset for predicting superconducting properties.

Therefore, we limit our analysis to the 18 electronic bands around to the Fermi surface. Each band is mapped onto a 32 $\times$ 32 $\times$ 32 grid, yielding band data with dimensions of 18 $\times$ 32 $\times$ 32 $\times$ 32. This targeted approach ensures that our dataset captures the most relevant features for predicting superconducting properties efficiently.

These band structure data, combined with experimentally reported $T_c$, form the basis of a ML training set. The dataset is stored in HDF5 format \href{https://www.hdfgroup.org/}{https://www.hdfgroup.org/}, providing a platform-independent, efficient means of accessing scientific and engineering data. In addition to the normalized band structure data, we also include several critical features for ML: orbital-resolved DOS data, chemical formulas, space group symmetries, lattice constants, atomic species, and atomic positions. These are depicted in Fig. \ref{fig1}.

\section{Technical Validation}

\begin{figure}[]
	\centering
	\includegraphics[width=0.45\textwidth]{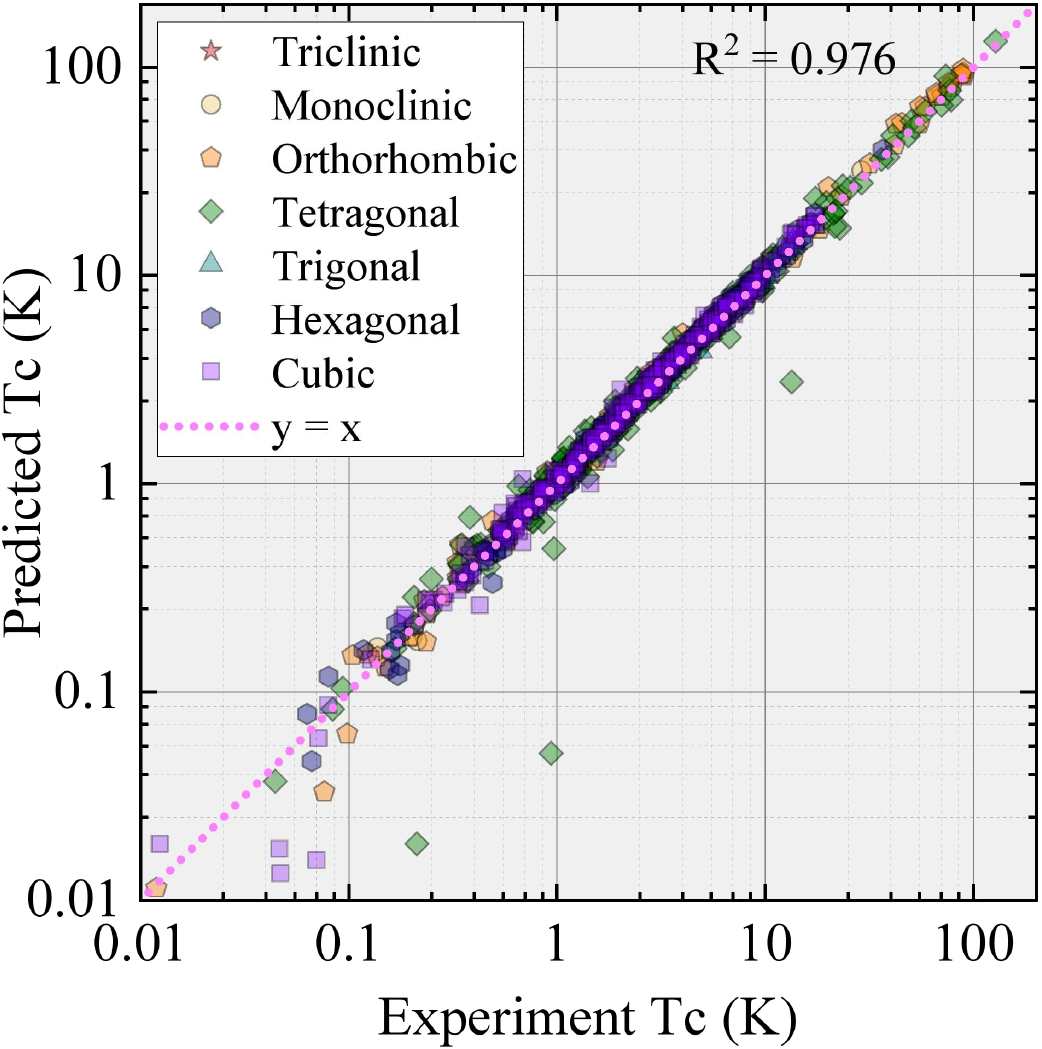}
	\caption{\label{fig6}Comparison between the deep learning model predicted $T_c$ and experimentally measured $T_c$ for the training set. The deep learning training process is as same as in ref\cite{Li2024}, but without considering data augmentation or cross-validation. }
\end{figure}
Fig. \ref{cif5} presents the energy band data for three representative superconductors in Superband, BCS superconductor MgB$_2$ with a hexagonal system \cite{Bohnen2001}, cuperate superconductor YBa$_2$Cu$_3$O$_7$ with an orthorhombic system\cite{Collocott1988}, and iron-based superconductor KFe$_2$Se$_2$ with a tetragonal system \cite{Ying2012}. The electronic band structures at high-symmetry points, along with the DOS plots, facilitate a more intuitive understanding of the distribution of electronic bands near the Fermi surface for researchers. The Fermi surface structure plots provide comprehensive visualizations of features closely associated with superconductivity, such as Fermi surface nesting. The electronic band data in training set demonstrates that our method for handling k-space preserves crystalline symmetry, even in hexagonal systems. 

For the technical validation and initial training of this dataset, we employed the 3D-Vision Transformer model \cite{dosovitskiy2020vit,Li2024} and compare the predicted $T_c$ with the experimental values. The goodness of fit between the predicted and experimental $T_c$ values is quantified using the coefficient of determination, $R^2$, defined by:
\begin{equation}\label{eq-r2}
R^2=1-\frac{S_{\text {Res}}}{S_{\text {Tot}}}=1-\frac{\sum_i\left(T_i-\hat{T}_i\right)^2}{\sum_i\left(T_i-\bar{T}\right)^2}, 
\end{equation}
where $T_i$ represents the predicted $T_c$ values, 
$\hat{T}_i$ denotes the average of predicted $T_c$ values, and $\bar{T}$ is the average experimental $T_c$.  The deep learning model's predictions, illustrated in Fig. \ref{fig6},  provide good agreement with the experimental superconductors, giving an $R^2=0.976$. The training outcomes validate the high quality and consistency of the data set in this study.

\section{Usage Notes}

The training dataset is made available in HDF5 format, and we recommend utilizing the h5py package for reading the file. Each read operation retrieves information on a specific superconductor, including attributes such as the chemical formula, $T_c$, space group, crystal system, and unit cell volume. The dataset directories contain detailed information on atomic positions, lattice structures, DOS, electronic bands, and Fermi surface data. Additionally, Python code is provided as a demonstration of how to access and read this file format.

For ease of use, a CSV file is included, which contains superconducting-related data, along with corresponding CIF files for the crystal structures. Most of these CIF files are sourced from the MP and the OQMD.

The Superband website \href{http://www.superband.work/}{http://www.superband.work/} offers a visual interface for exploring lattice structures, electronic band structures, and Fermi surface plots, providing an intuitive means of understanding the electronic properties of superconductors.

\nocite{*}
\bibliography{cite}
\end{document}